\documentclass[a4paper]{jpconf}
\usepackage{graphicx}
\usepackage{cite}
\usepackage{amsmath}
\usepackage{amssymb}
\usepackage{calrsfs}

\newcommand{\gae}{\lower 2pt \hbox{$\, \buildrel {\scriptstyle >}\over {\scriptstyle
\sim}\,$}}

\newcommand{\be}{\begin{equation}}
\newcommand{\ee}{\end{equation}}

\newcommand{\msum}{{\sum_{m=0}^\infty {}^{'}}}
\newcommand{\qsum}{\sum_{q=p,s}}

\newcommand{\mF}{\mathcal{F}}
\newcommand{\mS}{\mathcal{S}}
\newcommand{\tG}{\tilde\Gamma}

\newcommand{\kB}{k_\mathrm{B}}
\newcommand{\eb}{\bar{\epsilon}}
\newcommand{\omp}{\omega_\mathrm{p}}
\newcommand{\Li}{\mathrm{Li}}
\newcommand{\cth}{c_{\scriptstyle{\frac{3}{2}}}}
\newcommand{\fth}{{\scriptstyle{\frac{3}{2}}}}
\newcommand{\order}{\mathcal{O}}
\newcommand{\sigep}{4\pi\sigma}

\begin{document}
\title{Low temperature Casimir-Lifshitz free energy and entropy: the case of poor conductors}

\author{Simen A. \AA dn\o y Ellingsen$^1$, Iver Brevik$^1$, Johan S. H\o ye$^2$, Kimball A. Milton$^3$}

\address{$^1$ Department of Energy and Process Engineering, Norwegian University of Science and Technology, N-7491 Trondheim, Norway}
\address{$^2$Department of Physics, Norwegian University of Science and 
Technology, N-7491 Trondheim, Norway}
\address{$^3$Oklahoma Center for High Energy Physics and Department of Physics 
and Astronomy, The University of Oklahoma, Norman, OK 73019, USA}

\ead{simen.a.ellingsen@ntnu.no}

\begin{abstract}
The controversy concerning the temperature correction to the Casimir force has been ongoing for almost a decade with no view to a solution and has recently been extended to include semiconducting materials. We review some theoretical aspects of formal violations of Nernst's heat theorem in the context of Casimir Lifshitz thermodynamics and the role of the exponent of the leading term of the dielectric permittivity with respect to imaginary frequency. A general formalism for calculating the temperature corrections to free energy at low temperatures is developed for systems which do not exhibit such anomalies, and the low temperature behaviour of the free energy in a gap between half-spaces of poorly conducting materials modelled with a Drude type permittivity is calculated. 
\end{abstract}

\section{Introduction}

The Casimir force \cite{casimir48}, once merely a theoretical curiosity, is becoming the center of widespread attention in the wake of rapid developments in microtechnology. The enormous experimental progress made over the last decade towards accurately measuring this force 
\cite{lamoreaux97,mohideen98,roy99,harris00,ederth00,chan01,chen02,bressi02,decca05,bezerra06, chen07} has created the need to calculate the Casimir force with high accuracy in realistic settings, taking into account such effects as material optical properties, surface roughness and geometry effects. Reviews of recent progress include \cite{milton04,buhmann07}.

It was realised quite early that an ambiguity existed as to the interpretation of Lifshitz' formula \cite{lifshitz55} for the Casimir attraction between dielectric half-spaces: when describing an ideal metal by taking the permittivity to infinity, different results were obtained at finite temperatures depending on the way the limit was taken. The ambiguity was originally sidestepped by prescription \cite{schwinger78} and re-examined only much later by Bostr\"om and Sernelius \cite{bostrom00} who concluded that due to the finite relaxation time of conduction electrons in a metal the transverse electric (TE) reflection coefficient of a metal-vacuum interface must vanish in the zero frequency limit, contrary to Casimir's ideal metal approximation in which reflection coefficients are set to unity at all energies. This was further supported by another study by H\o ye et al.\ \cite{hoye03}.
For finite temperatures this vanishing of the TE zero frequency reflection coefficient leads to a prediction of a relatively large reduction of the Casimir force between metal plates at finite temperature, up to 15\% at 300K. The Bostr\"om-Sernelius analysis was opposed on thermodynamical grounds for violating Nernst's theorem (the third law of thermodynamics) which states that a nondegenerate system must have zero entropy at zero temperature\cite{klimchitskaya01}. Moreover it was concluded that the series of high accuracy experiments at Purdue exclude the thermal correction predicted by the theory in which the TE zero frequency mode does not contribute \cite{decca05, bezerra06}. The debate is summarised in recent reviews \cite{brevik06, klimchitskaya06}.

Recently an analogous ambiguity in the Lifshitz formalism was brought up for the case of semiconductors \cite{geyer05, klimchitskaya06b, geyer06, klimchitskaya08}. A formal violation of Nernst's theorem is once again the difficulty, this time due to discontinuous behaviour in the transverse magnetic (TM) reflection coefficient, whose value in the limit of zero frequency depends intimitely on the way the small density of conducting electrons in semiconducting materials at finite temperatures are taken into account.

However, we do not expect such formal violations of Nernst's theorem stemming from the mathematical subtleties of the Lifshitz formula to have implications for the physics of the problem. In \cite{hoye03} it was concluded that on physical grounds, no TE zero mode should be present for real metals, and recently a quantum statistical mechanical treatment came to the same result \cite{buenzli08}.
For semiconductors, earlier statistical mechanical analyses by Jancovici and \v{S}amaj \cite{jancovici04,jancovici05} and by Buenzli and Martin \cite{buenzli05} for ionic systems are of interest. What is found is that the effective separation between the plates increases as twice the ionic shielding length, which implies a non-local behaviour of the dielectric function. This increase in effective separation also means that the ionic contribution to reflectivity vanishes with vanishing ionic concentration. The results of refs.\ \cite{jancovici04, jancovici05, buenzli05} are restricted to ionic systems, but in the high temperature (classical) limit they recover the ideal metal result corresponding to no TE zero mode. 

\section{Formal violations of Nernst's heat theorem; general theory}

The Lifshitz formula expresses the Casimir free energy between parallel surfaces described by polarisation specific reflection coefficients $r_q$ where $q\in \{p,s\}$ is the polarisation (assuming specular reflection and no coupling between $p$ and $s$ modes). For simplicity we shall assume the surfaces to be identical in the following, in which case the Casimir free energy at temperature $T$ reads
\be\label{F}
  \mF(a) = \frac{T}{2\pi} \msum \int_{\zeta_m}^\infty \rmd \kappa\kappa \qsum \ln(1-r_q^2 \rme^{-2\kappa a})
\ee
wherein $a$ is the plate separation, $p,s$ denotes TM and TE polarisations respectively, and $\rmi\zeta_m$ are the (imaginary) Matsubara frequencies so that $\zeta_m = 2\pi mT$. As conventional, the prime on the summation mark signifies that the $m=0$ term be taken with half weight. We will be using natural units $\kB=\hbar=c=1$ throughout. Henceforth we will frequently omit the subscript $m$ on $\zeta_m$ and the various quantities depending on it. The integral in (\ref{F}) is over all transverse momenta $\mathbf{k}_\perp$ of the field ($\perp$ denotes a direction parallel to the surfaces) and the substitution $\kappa^2=\mathbf{k}_\perp^2+\zeta^2$ has been made.

In the case where the interfaces are between vacuum and a half-space made of dielectric material, the Fresnel reflection coefficients read
\be
  r_s = \frac{\kappa-\tilde\kappa}{\kappa+\tilde\kappa};~~~r_p = \frac{\varepsilon\kappa-\tilde\kappa}{\varepsilon\kappa+\tilde\kappa};~~~ \tilde\kappa \equiv \sqrt{\kappa^2 + \zeta^2(\varepsilon-1)}.
\ee
Here $\varepsilon=\varepsilon(\rmi\zeta)$ denotes the dielectric permittivity relative to vacuum. 

It is straightforward to verify that the values of $r_q$ in the limit $\zeta\to 0$ depend on the leading exponent of $\varepsilon(\rmi \zeta)$ as this limit is approached. For materials with mobile charges, models of the permittivity will typically diverge in the zero frequency limit, whereas that of a pure dielectric isolator reaches a finite value, $\lim_{\zeta\to 0}\varepsilon(\rmi\zeta) = \eb$. Assuming $\varepsilon(\rmi\zeta)\sim (\zeta/\tilde\omega)^\lambda$ ($\tilde\omega$ is a constant) as $\zeta\to 0$, one readily obtains the limits arrayed in table \ref{table} in which 
\be\label{trs}
  \tilde r_s(\kappa)=-\frac{\kappa^2}{\tilde\omega^2}\left(\sqrt{1+\frac{\tilde\omega^2}{\kappa^2}}-1\right)^2\leq 0.
\ee

\begin{center}
  \begin{table}[h]
    \begin{minipage}{5pc}
      \begin{tabular}{lcc}
        \br
        $\lambda$\hspace{20pt}&$r_s$ &$r_p$\\
        \mr
        $0$& $0$& $\frac{\eb-1}{\eb+1} $\\
        $-1$& $0$&1\\
        $-2$& $\tilde r_s(\kappa)$&1\\
        $< -2$ &$ -1$ & 1\\
        \br
      \end{tabular}
    \end{minipage}
    \hspace{5pc}
    \begin{minipage}[b]{15pc}
      \caption{\label{table}Values of $r_q(\rmi\zeta\to 0, \kappa)$ for different exponents $\lambda$.} 
    \end{minipage}
  \end{table}
\end{center}

The model permittivities which cause formal violation of Nernst's theorem have the common trait that the exponent $\lambda$ takes one value at all finite temperatures which changes abruptly at exactly $T=0$. A general treatment demonstrates that such temperature dependence is necessary in order for a formal violation of the theorem to occur \cite{ellingsen08,intravaia08} as we will briefly explain.

An example is the application of a Drude model to describe the dielectric response of an infinitely large and perfectly pure metal lattice, for which  $\varepsilon(\rmi\zeta)$ is modelled as
\be
  \varepsilon(\rmi\zeta) = 1 + \frac{\omp^2}{\zeta[\zeta+\nu(T)]}.
\ee
For a real metal sample of finite size, $\nu(T)$ reaches a nonzero value at zero temperature due to electron scattering on boundaries, impurities and imperfections, and $\lambda=-1$ for all temperatures. In this case entropy vanishes in the zero temperature limit as it should \cite{hoye07, brevik08}. In a perfect lattice of infinite size, however, electron relaxation is solely due to scattering on thermal phonons, so that $\nu\sim T^5$ as $T\to 0$. Thus $\lambda$ changes from $-1$ to $-2$ at $T=0$, making $r_s$ jump discontinuously from zero to a finite value as seen in table \ref{table}. Clearly $\tilde\omega= \omp$ in (\ref{trs}) in the case $\lambda=-2$.

Another example is the semiconductor whose conductivity vanishes as a function of $T$. If a Drude model is used to model the permittivity of such a material,
\be\label{epSemi}
  \varepsilon(\rmi\zeta) = 1 + \frac{\eb-1}{1+\zeta^2/\omega_0^2} + \frac{4\pi\sigma(T)}{\zeta},
\ee
a formal violation occurs when $\sigma(T)$ vanishes at exactly $T=0$. In this case $\lambda=-1$ at all $T$ until absolute zero, where it skips to $\lambda=0$ and the magnitude of $r_p$ jumps discontinuously. 

Let 
\be
  \lim_{\zeta\to 0}r_q = R_q(\kappa;\lambda)
\ee
as tabulated (note that apart from the $\lambda=-2$ $s$-mode, $R_q$ is independent of $\kappa$). It can be shown that when reflection coefficients jump discontinuously at $\zeta=0$, the free energy obtains a term linear in temperature equal to the difference of the $m=0$ terms of (\ref{F}) as obtained with the two zero-frequency reflection coefficients respectively. If the leading $\zeta$ exponent of $\varepsilon$ changes from $\lambda_1$ to $\lambda_2$ at exactly $T=0$, therefore, it leads to a residual entropy $\mS = -\partial \mF/\partial T$ at zero temperature
\be\label{Sgeneral}
  \mS_{\lambda_1\to \lambda_2} = \frac{1}{4\pi}\qsum\int_0^\infty \rmd \kappa\kappa \ln \frac{1-R^2_q(\kappa; \lambda_2)\rme^{-2\kappa a}}{1-R^2_q(\kappa; \lambda_1)\rme^{-2\kappa a}}.
\ee

In particular, when $R_q(\kappa;\lambda_1)=R_q(\lambda_1)$ and $R_q(\kappa;\lambda_2)=R_q(\lambda_2)$ one may use the relation
\be
  \int_0^\infty \rmd\kappa\kappa\ln[1-R^2\rme^{-2\kappa a}] = -\frac{1}{4a^2}\Li_3(R^2)
\ee
where $\Li_n(x)$ is the $n$th order polylogarithmic function
\be\label{polylog}
  \Li_n(x) = \sum_{l=1}^\infty \frac{x^l}{l^n}
\ee
to write
\be\label{resEntr}
  \mS_{\lambda_1\to \lambda_2} = \frac{1}{16\pi a^2}\qsum\left\{\Li_3[R_q^2(\lambda_1)]-\Li_3[R_q^2(\lambda_2)]\right\}.
\ee

In the particular cases of Drude modelled metals and semiconductors discussed above it follows immediately from table \ref{table} and Eqs.\ (\ref{Sgeneral}) and (\ref{resEntr}) that, respectively,
\begin{subequations}
  \begin{align}
    \mS_{-1\to -2} &= \frac{1}{4\pi}\int_0^\infty \rmd \kappa\kappa \ln [1-\tilde r^2_s(\kappa)\rme^{-2\kappa a}];\label{Smetal} &\mbox{(metals)}\\
    \mS_{-1\to 0} &= \frac{1}{16\pi a^2}\left\{\zeta(3)- \Li_3[R_p^2(0)]\right\}&\mbox{(semiconductors)}
  \end{align}
\end{subequations}
where $R_p(0) = (\eb-1)/(\eb+1)$ 
and where we have used $\Li_n(1)=\zeta(n)$, the Riemann zeta function. These two exponent transitions are those which come into play for metals and dielectrics respectively, or more precisely, upon plugging a Drude-type permittivity model with vanishing $\nu(T)$ or $\sigma(T)$ into the Lifshitz formula and extrapolating to zero temperature. Other exponent transitions, naturally, would give other zero point entropy expressions.

Note that by letting $\tilde\omega\to \infty$ in (\ref{trs}) so that $\tilde r_s(\kappa) \to -1$, the entropy (\ref{Smetal}) becomes
\be
  \mS_{-1\to -2} \buildrel{\tilde\omega\to \infty}\over{\longrightarrow} -\frac{\zeta(3)}{16\pi a^2},\\
\ee
which is the well known result for the so-called modified ideal metal model obtained by this procedure \cite{hoye03}.

\section{Free energy temperature correction for poor Drude conductor}

The model for the conductivity we will be studying in the following is assumed not to depend on $T$ within a finite range of temperatures including $T=0$, in which case it is clear from the above that Nernst's theorem will be satisfied. While the consideration of such a model cannot resolve such anomalies as reported in the previous section, it is nonetheless useful to establish benchmark results in various models within the Lifshitz formalism which exhibits very nontrivial behaviour in the joint limit of zero temperature and frequency $\zeta$.

We consider a semiconductor modelled by a Drude type permittivity such as (\ref{epSemi}), but where we assume $\sigma$ to be constant within a finite range of low temperatures including $T=0$ (the case $\sigma=0$ was worked out in \cite{geyer05}). A more detailed treatment of this model may be found in \cite{ellingsen08b}. 

The method used is to note that $\mF(a,T)$ depends on $T$ only through the prefactor $T$ and the Matsubara frequencies $\zeta_m=2\pi mT$, so we may write the free energy (\ref{F}) on the form\footnote{Note that this convention differs sightly from that of \cite{ellingsen08b}.}
\be\label{Fg}
  \mF = T\qsum f_q(a)\msum g(\mu)
\ee
where the function $f_q(a)$ is a convenient prefactor and we use the shorthand notation $\mu\equiv mt$ where $t=T/T_0$ is a dimensionless rescaled temperature to be defined in Eq.\ (\ref{t}) below. 

In the limit $T=0$ the sum becomes an integral, and we are interested in the difference between sum and integral, which may be found from the Euler-Maclaurin formula. It turns out $g(\mu)$ is not analytical at $m=0$ so it is necessary to start the sum at $m=1$ (or a higher value), writing
\begin{align}
  \tG &\equiv \left[{\sum_{m=0}^\infty}'-\int_0^\infty \rmd m\right]g(mt) = \frac{1}{2}g(0)
-\int_0^1g(mt)\rmd m +\frac12g(t)- \sum_{k=1}^\infty \frac{B_{2k}}{(2k)!}g^{(2k-1)}(t)\notag \\
    &=\frac{1}{2}g(0)-\int_0^1g(mt)\rmd m+\frac{1}{2}g(t)-\frac{1}{12}g'(t)
    +\frac{1}{720}g'''(t)-\dots.\label{EM}
\end{align}
Since the correction terms in (\ref{EM}) are evaluated at $m=1$ and we are considering small $T$, we may choose $T_0\gg T$ so that $\mu\ll 1$. We anticipate that when expanded for small $\mu$, the function $g(\mu)$ is of the form
\be\label{gExp}
  g(\mu)\sim c_0 + c_1 \mu + \cth \mu^\frac{3}{2}+ c_{2l}\mu^2 \ln(\mu) +c_2\mu^2 + c_3\mu^3+..., ~~\mu\to 0
\ee
Upon insertion into (\ref{EM}) one finds that the $c_0$ and $c_2$ terms do not contribute. Terms of integer powers of $\mu$ are determined by a finite number of terms in the series (\ref{EM}), but for the terms containing logarithms or half-integer powers every term contributes. The series obtained is asymptotic but a meaningful value may nonetheless be assigned to all terms of (\ref{gExp}) by defining the series by Borel summation\footnote{See appendix of \cite{ellingsen08b}} or zeta regularisation. The result for polarisation mode $q=p,s$ is \cite{ellingsen08b}
\be\label{dFexp}
  \Delta\mF_q = Tf_q(a)\tG_q = Tf_q(a)\left[-\frac{c_1}{12}t+ \zeta(-\fth)\cth t^\frac{3}{2}+\frac{\zeta(3)c_{2l}}{4\pi^2}t^2+\frac{c_3}{120}t^3+...\right]_q
\ee
giving terms proportional to respectively $T^2, T^\frac{5}{2}, T^3$ and $T^4$. It is understood that the coefficients $c_1$ through $c_3$ are polarisation mode specific. We shall content ourselves with expanding free energy to order $T^3$ in the present paper. 

Equation (\ref{dFexp}) effectively reduces the problem to one of determining the expansion coefficients of Eq.\ (\ref{gExp}). This task is still not trivial, however, and we consider only the ``intermediate asymptotic'' region in which conductivity is very small compared to inverse separation but much greater than temperature
\be\label{asymptote}
  T \ll 4\pi \sigma \ll \frac{1}{a}.
\ee

It is convenient now to define $T_0=2\sigma$, that is 
\be\label{t}
  t = \frac{T}{T_0}\equiv \frac{\zeta_1}{4\pi\sigma} = \frac{2\pi T}{4\pi\sigma}.
\ee
Assumption (\ref{asymptote}) ensures that $t\ll 1$. 
The frequency which enters into (\ref{EM}) is $\zeta_1\sim T \ll \sigma$, so $\varepsilon(\rmi \zeta)$ simplifies to
\be\label{ep}
  \varepsilon = \eb + \frac1{\mu}
\ee
We will consider the $p$ and $s$ modes individually in the following. 

\section{The TM mode}

The TM mode expression for Casimir Lifshitz free energy exhibits highly nontrivial behaviour near zero temperature. To simplify matters we note from (\ref{asymptote}) that the quantity
\be
  \alpha \equiv 2a(4\pi\sigma)
\ee
obeys $\alpha\ll 1$. We will determine free energy corrections perturbatively in powers of $\alpha$ in order to obtain analytical results for the coefficients in (\ref{gExp}).

Substituting the integration variable 
\be
  x = 2\kappa a =\frac{\kappa\alpha\mu}{\zeta}
\ee
the TM free energy may be written
\be\label{xLif}
  \mF_p = \frac{(\sigep)^3t}{4\pi^2\alpha^2}\msum\left\{ \int_{\alpha\mu}^\infty \rmd xx\ln(1-r_p^2 \rme^{-x})\right\}\equiv \frac{(\sigep)^3t}{4\pi^2\alpha^2}\msum g_p(m)
\ee
where $g_p(m)$ is now chosen to be the expression within the curly braces. 
To leading order in $\alpha$ one finds
\begin{subequations}
\begin{align}
  \ln(1-r_p^2 \rme^{-x}) &= \ln(1-A_\mu \rme^{-x}) + \order(\alpha^2); \\
  A_\mu &\equiv \left(\frac{1+(\eb-1)\mu}{1+(\eb+1)\mu}\right)^2.
\end{align}
\end{subequations}
With the reflection squared coefficient now a constant with respect to $x$ the integral in (\ref{xLif}) can be solved explicitly and expansion in $\mu$ as in (\ref{gExp}) yields the coefficients $c_1$ and $c_{2l}$ to leading order in $\alpha$ as simply \cite{ellingsen08b}
\be
   c_1 = \frac{2\pi^2}{3};~~~c_{2l} = 8
\ee
and with (\ref{dFexp}) the low temperature correction to free energy is to leading order in $\alpha$ in terms of $T$ and $\sigma$ reads:
\be\label{dFTM}
  \Delta\mF_p = -\frac{\pi^2T^2}{72(\sigep)a^2} 
  + \frac{\zeta(3)T^3}{\pi(\sigep)^2a^2} .
\ee
An examination shows that the correction to this result is approximately a factor $\alpha^2$ smaller as anticipated.

\section{The TE mode}

A similar procedure is performed for the $s$ (or TE) mode. With the substitution
\be
    x = \frac{\kappa}{\zeta\sqrt{\varepsilon(i\zeta)-1}} = \frac{\kappa \mu}{\chi \zeta}; ~~~\chi \equiv \sqrt{\mu + (\eb-1)\mu^2},
\ee
the TE term of the free energy (\ref{F}) may be written
\be\label{Fs}
  \mF_s = \frac{(\sigep)^3t}{4\pi^2} \msum \left\{ \chi^2\int_{\mu/\chi}^\infty \rmd x x\ln(1-r_s^2 \rme^{-\alpha\chi x})\right\}=\frac{(\sigep)^3t}{4\pi^2} \msum g_s(m).
\ee
Again $g_s(m)$ is defined as the expression between the curly braces and in the new variables the squared reflection coefficient reads
\be
  r^2_s(x) = (\sqrt{x^2+1}-x)^4.
\ee

Expanding the logarithm of (\ref{Fs}) to linear order in $\alpha$,
\be \label{gsExp}
  \ln(1-r_s^2 \rme^{-\alpha\chi x}) = \ln(1-r_s^2) + \frac{\alpha\chi x r_s^2}{1-r_s^2} + \order (\alpha^2)
\ee
allows us to determine the temperature corrections for these orders of $\alpha$. Both terms on the right hand side of (\ref{gsExp}) give integrals over $x$ which are explicitly solvable, and subsequent expansions in powers of $\mu$ give the coefficients
\be
  c_1 = -\frac{1}{4}(2\ln 2 - 1); ~~~ c_{2l} = -\frac{1}{4}; ~~~ \cth = \frac{\alpha}{12}.
\ee
Clearly $\cth\ll c_1, c_{2l}$, being of linear order in $\alpha$. The temperature corrections for the TE mode thus read
\be\label{dFTE}
  \Delta \mF_s = \frac{(\sigep)T^2}{48}(2\ln 2-1) + \frac1{6}\sqrt{2\pi}\zeta(-\fth)a(\sigep)^\frac{3}{2}T^\frac{5}{2} - \frac{\zeta(3)T^3}{8\pi}+\order(T^\frac{7}{2}).
\ee
This result is in fact in perfect agreement with that obtained for Drude metals \cite{hoye07} but includes one more order. Thus the concordance implies that a further expansion in $\alpha$ would only yield corrections of higher orders in temperatures, and that for the case of the TE mode the expansion in $\alpha$ was not essential for obtaining this result. Eq.\ (\ref{dFTE}) is therefore valid also when $\alpha$ is not small, as is the case for a good conductor. 

Notably the temperature corrections are independent of $\eb$ to order $T^3$ for both polarisations and only enters in higher orders, an insight not included in \cite{hoye07}.

\section{Numerical verification of results}

\begin{figure}[htb]
\begin{minipage}{18pc}
\includegraphics[width=18pc]{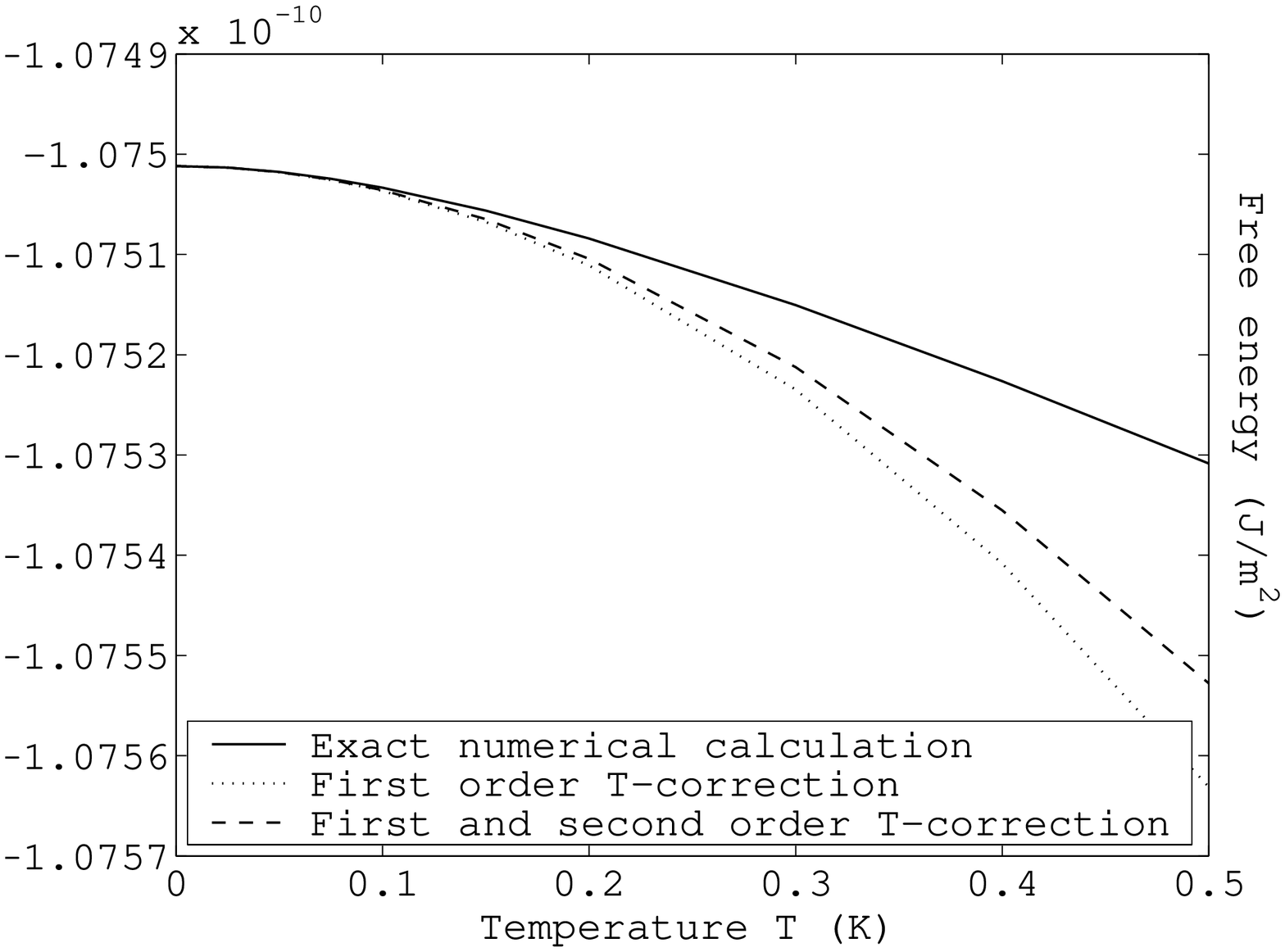}
\caption{\label{fig_FTM}Free energy $\mF_p$ calculated using direct numerical calculation and the theoretical correction (\ref{dFTM}) shifted to coincide with the numerical result at $T=0$.}
\end{minipage}\hspace{2pc}%
\begin{minipage}{18pc}
\includegraphics[width=18pc]{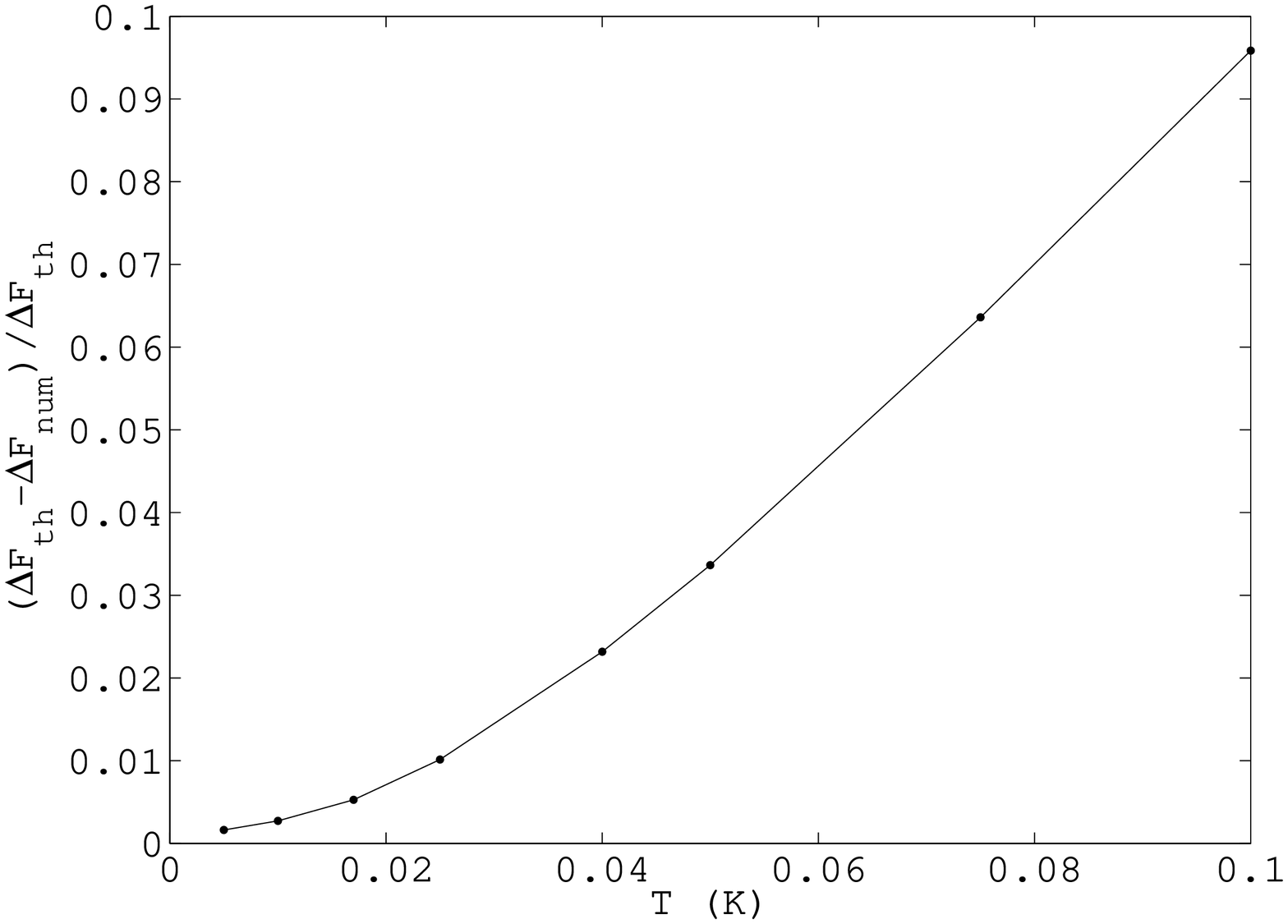}
\caption{\label{fig_R}Plot of the quantity $R$ from Eq.\ (\ref{R}). }
\end{minipage} 
\end{figure}

The final results (\ref{dFTM}) and (\ref{dFTE}) have been checked numerically to verify their correctness \cite{ellingsen08b}. A plot of the theoretical correction (\ref{dFTM}) compared to a direct numerical calculation of the free energy is provided in figure \ref{fig_FTM}. 

A much more sensitive test is provided by the quantity
\be\label{R}
  R_p = \frac{\Delta \mF_p^\text{th}-\Delta \mF_p^\text{num}}{\Delta \mF_p^\text{th}}
\ee
where $\Delta \mF_p^\text{th}$ is the theoretically predicted free energy correction (\ref{dFTM}) and $\Delta \mF_p^\text{num}$ is that found by direct numerical calculation. The data used for calculation are (in SI units) $a=1000$nm, $\sigma^\text{SI}/\epsilon_0=10^{12}$s$^{-1}$, $\eb=11.66$ and $\omega_0=8\cdot 10^{15}$s$^{-1}$.

We have found that $\Delta \mF_p^\text{th}$ is of the form $\Delta \mF_p^\text{th} = -CT^2(1-C_1 T)$ and assume $\Delta \mF_p^\text{th}$ to have the expansion
\be
  \Delta \mF_p^\text{num} = -DT^2(1-D_1T+D_2T^2+\dots),
\ee
which predicts the following expansion for $R$:
\be
  R = \frac{C-D}{C} -\frac{D}{C}(C_1-D_1)T -\frac{D}{C}[D_2+C_1(C_1-D_1)]T^2 + \dots.
\ee
In the special case where $C=D$ and $C_1=D_1$, this becomes
\be\label{Rcorrect}
  R = -D_2T^2 + \mathcal{O}(T^3).
\ee
Thus if the coefficient $C$ is incorrect $R$ would not converge to $0$ at $T=0$, and an incorrect $C_1$ would show as a linear behaviour at small temperatures. None of these effects are perceptible in the figure, demonstrating that the corrections to (\ref{dFTM}) are small as predicted.

\begin{figure}[htb]
\begin{minipage}{18pc}
\includegraphics[width=18pc]{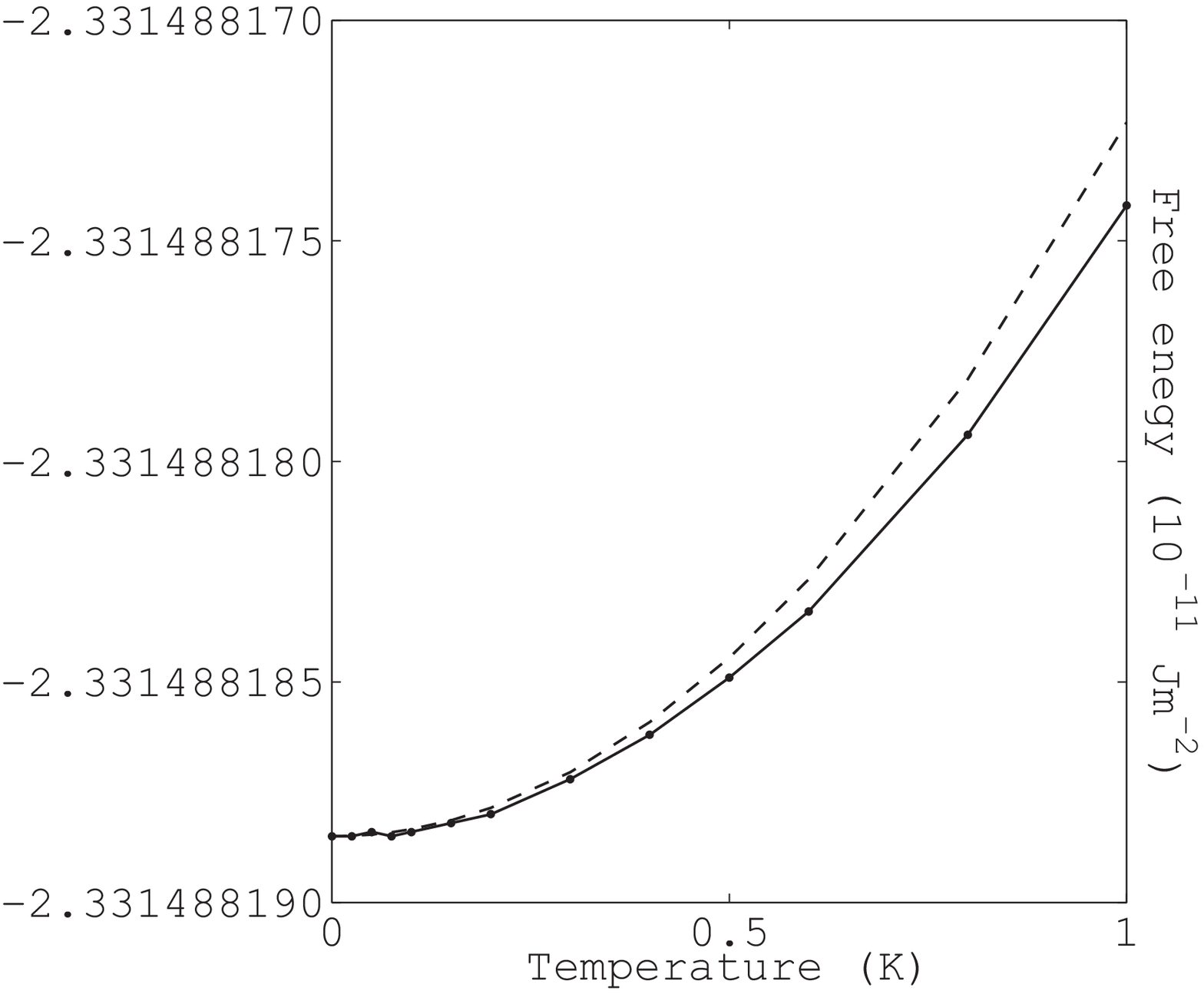}
\caption{\label{fig_FTE}Free energy $\mF_s$ calculated using direct numerical calculation and the theoretical correction (\ref{dFTE}) to order $T^2$ shifted to coincide with the numerical result at $T=0$.}
\end{minipage}\hspace{2pc}%
\begin{minipage}{18pc}
\includegraphics[width=18pc]{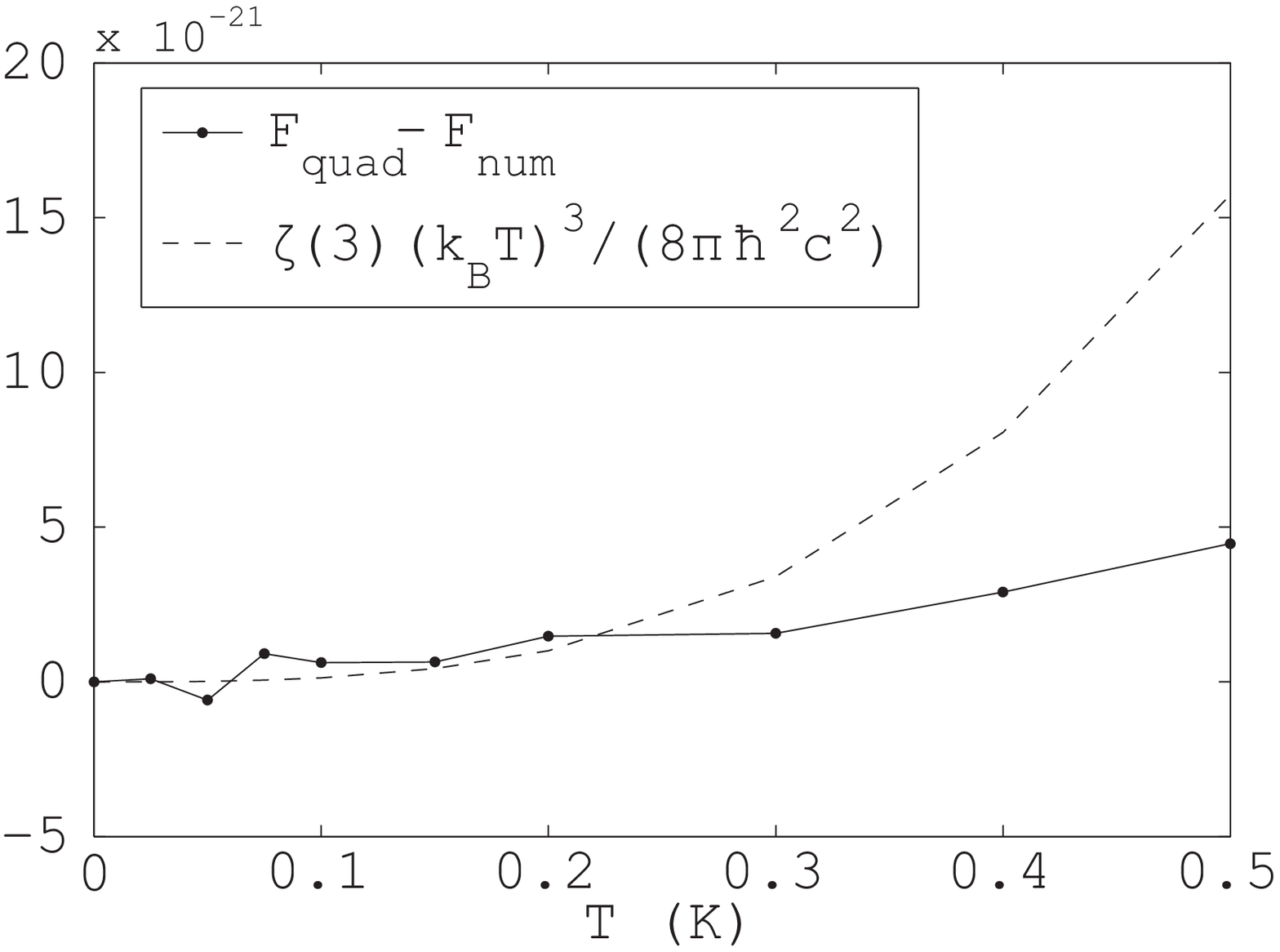}
\caption{\label{fig_diffTE}The difference between the graphs in figure \ref{fig_FTE} plotted against the $T^3$ term of (\ref{dFTE}).}
\end{minipage} 
\end{figure}

A similar high precision check of the result for the $s$ mode (\ref{dFTE}) was not possible with these numerical parameter values because the correction relative to the free energy at zero temperature is extremely small, of order $10^{-9}$ at $1$K. The correctness of the terms proportional to $T^2$ and $T^\frac{5}{2}$ was however thoroughly verified for a good conductor in \cite{hoye07}. The term $\propto T^3$ is numerically elusive because while requiring very high accuracy for verification when $\alpha\ll 1$, it is completely dominated by other terms for good conductors. A comparison of a direct numerical calculation with the prediction (\ref{dFTE}) to order $T^2$ in figure \ref{fig_FTE} reveals that the difference between these graphs, plotted in figure \ref{fig_diffTE}, is in the same order of magnitude as the $T^3$ term in (\ref{dFTE}), while the term proportional to $T^\frac{5}{2}$ is too small to be visible at this level.

\section{Conclusions}

We have reviewed the theory of formal violations of Nernst's heat theorem emphasising the way such a formal violation can only occur when the leading order behaviour of $\varepsilon(\rmi \zeta)$ with respect to $\zeta$ undergoes a discontinuous change at exactly $T=0$. Such apparent problems with the Lifshitz formalism occur when the double limit where $T$ and $\zeta$ are both taken to zero is not unique and depends intimately on the exact way in which a material's dielectric (and, in general, magnetic) response is modelled in this limit. As a general remark, Nernst's theorem concerns zero temperature only, and it is not \emph{a priori} clear that one can simply extrapolate between these two very different temperatures and use a result at one temperature to draw conclusions at another. In particular, if a system behaves essentially different at $T=0$ than at room temperature, a formal violation of Nernst's theorem by extrapolation is not necessarily worrisome. The results obtained complement those found in the case of zero conductivity in \cite{geyer05} and generalise TE mode calculations for Drude metals in \cite{hoye07}.

Using a Drude type model to describe a poor conductor whose conductivity stays finite at zero temperature we establish the low temperature corrections to Casimir Lifshitz free energy between two identical half-spaces separated by vacuum. As modelled, both TE and TM modes exhibit a quadratic temperature behaviour at low temperatures.


\end{document}